\documentclass[nojss]{jss}
\usepackage[latin1]{inputenc}
\usepackage{amsmath}
\usepackage{amsfonts}
\usepackage{amssymb}
\usepackage{graphicx}
\usepackage{color}

\def\vect#1{{\vec{#1}}}                               
\def\mat#1{{\mathbf{#1}}}	                      
\def\dotP#1#2{\vect{#1}\cdot\vect{#2}}		      
\def\partderiv#1#2{\frac{\partial{}#1}{\partial{}#2}} 

\def\prob#1{\Prob\left(#1\right)}		      

\author{Asad Hasan\\ Scientific Computing \\ Sentrana Inc. \And
        Wang Zhiyu\\ Dept of Mathematical Sciences \\ Carnegie Mellon University \And
        Alireza S. Mahani\\ Scientific Computing \\ Sentrana Inc.}
\title{Fast Estimation of Multinomial Logit Models: \\ \proglang{R} Package \pkg{mnlogit}}

\Plainauthor{Asad Hasan, Zhiyu Wang, Alireza S. Mahani} 
\Plaintitle{Fast Estimation of Multinomial Logit Models: R Package mnlogit} 
\Shorttitle{Fast Estimation of Multinomial Logit Models: \proglang{R} package \pkg{mnlogit}} 

\Abstract{ 
 We present \proglang{R} package \pkg{mnlogit} for training multinomial logistic regression models, particularly those involving a large number of classes and features.
 Compared to existing software, \pkg{mnlogit} offers speedups of 10x-50x for modestly sized problems and more than 100x for larger problems.
Running in parallel mode on a multicore machine gives upto 4x additional speedup on 8 processor cores.  
\pkg{mnlogit} achieves its computational efficiency by drastically speeding up computation of the log-likelihood function's Hessian matrix through exploiting structure in matrices that arise in intermediate calculations.
}
\Plainkeywords{logistic regression, multinomial logit, discrete choice, large scale, parallel, econometrics} 

\Address{
  Asad Hasan\\
  Scientific Computing Group\\
  Sentrana Inc.\\
  1725 I St NW\\
  Washington, DC 20006\\
  E-mail: \email{asad.hasan@sentrana.com}\\

  Zhiyu Wang\\
  Department of Mathematical Sciences\\
  Carnegie Mellon University\\
  5000 Forbes Ave\\
  Pittsburgh, PA 15213\\

  Alireza S. Mahani\\
  Scientific Computing Group\\
  Sentrana Inc.\\
  1725 I St NW\\
  Washington, DC 20006\\
  E-mail: \email{alireza.mahani@sentrana.com}\\
}

\begin{document}





\section{Introduction}
\label{section: introduction}
Multinomial logit regression models, the multiclass extension of binary logistic regression, have long been used in econometrics in the context of modeling discrete choice \citep{MCFAD:74, BHAT:95, TRAIN:03} and in machine learning as a linear classification technique~\citep{HastieTibBook} for tasks such as text classification~\citep{Nigam1999}.
Training these models presents the computational challenge of having to compute a large number of coefficients which increases linearly with the number of classes and the number of features.
Despite the potential for multinomial logit models to become computationally expensive to estimate, they have an intrinsic structure which can be exploited to dramatically speedup estimation. 
Our objective in this paper is twofold: first we describe how to exploit this structure to optimize computational efficiency, and second, to present an implementation of our ideas in our \proglang{R}~\citep{Rlang} package~\pkg{mnlogit} which is available from CRAN at: \url{http://cran.r-project.org/web/packages/mnlogit/index.html}.

An older method of dealing with the computational issues involved in estimating large scale multinomial logistic regressions has been to approximate it as a series of binary logistic regressions~\citep{BeggGray1984}. 
In fact the \proglang{R} package \pkg{mlogitBMA}~\citep{mlogitBMA} implements this idea as the first step in applying Bayesian model averaging to multinomial logit data.
Large scale logistic regressions can, in turn, be tackled by a number of advanced optimization algorithms \citep{KomarekMoore, LinLogistic2009}.
A number of recent \proglang{R} packages have focussed on slightly different aspects of estimating regularized multinomial logistic regressions. 
For example: package \pkg{glmnet}~\citep{glmnet} is optimized for obtaining the entire $L1$-regularized paths and uses the coordinate descent algorithm with `warm starts', package \pkg{maxent}~\citep{JurkaMaxent} is intended for large text classification problems which typically have very sparse data and the package \pkg{pmlr}~\citep{pmlr} which penalizes the likelihood function with the Jeffreys prior to reduce first order bias and works well for small to medium sized datasets.
There are also \proglang{R} packages which estimate plain (unregularized) multinomial regression models. 
Some examples are: the \pkg{VGAM} package~\citep{VGAM}, the \code{multinom} function in package \pkg{nnet}~\citep{nnet} and package the \pkg{mlogit}~\citep{mlogit}.

Of all the \proglang{R} packages previously described, \pkg{mlogit} is the most versatile in the sense that it handles many data types and extensions of multinomial logit models (such as nested logit, heteroskedastic logit, etc.). 
These are especially important in econometric applications, which are motivated by the utility maximization principle \citep{MCFAD:74}, where one encounters data which depends upon \emph{both} the observation instance and the choice class.
Our package \pkg{mnlogit} provides the ability of handling these general data types while adding the advantage of very quick computations.
This work is motivated by our own practical experience of the impossibility of being able to train large scale multinomial logit models using existing software.

In \pkg{mnlogit} we perform maximumum likelihood estimation (MLE) using the Newton-Raphson (NR) method.
We speed up the NR method by exploiting structure and sparsity in intermediate data matrices to achieve very fast computations of the Hessian of the log-likelihood function.   
This overcomes the NR method's well known weakness of incurring very high per-iteration cost, compared to algorithms from the quasi-Newton family \citep{Nocedal1992, Nocedal1990}.
Indeed classical NR estimations of multinomial logit models (usually of the Iteratively Reweighted Least Square family) have been slow for this very reason. 
On a single processor our methods have allowed us to achieve speedups of 10x-50x compared to \pkg{mlogit} on modest-sized problems while performing~\emph{identical} computations. 
In parallel mode\footnote{Requires \pkg{mnlogit} to be compiled with OpenMP support (usually present by default with most \proglang{R} installations, except on Mac OS X).}, \pkg{mnlogit} affords the user an additional speedup of 2x-4x while using up to 8 processor cores.

We provide a simple formula-based interface for specifiying a varied menu of models to \pkg{mnlogit}.  
Section~\ref{section: data format and model specification} illustrates aspects of the formula interface, the expected data format  and the precise interpretations of variables in \pkg{mnlogit}.
To make the fullest use of \pkg{mnlogit} we suggest that the user understand the simple \proglang{R} example worked out over the course of this section.  
Section~\ref{section: algorithms and optimization} and Appendix~\ref{appendix: log-likelihood differentiation } contain the details of our estimation procedure, emphasizing the ideas that underlie the computational efficiency we achieve in \pkg{mnlogit}.
In Section~\ref{section: benchmarking performance} we present the results of our numerical experiments in benchmarking and comparing \pkg{mnlogit}'s performance with other packages while Appendix~\ref{appendix: timing tests} has a synopsis of our timing methods. 
Finally Section~\ref{section: discussion} concludes with a short discussion and a promising idea for future work.


\section[On using mnlogit]{On using \pkg{mnlogit}}
\label{section: data format and model specification}

The data for multinomial logit models may vary with both the choice makers (`individuals') and the choices themselves.
Besides, the modeler may prefer model coefficients that may (or may not) depend on choices.  
In \pkg{mnlogit} we try to keep the user interface as minimal as possible without sacrificing flexibility.
We follow the interface of the \code{mlogit} function in package \pkg{mlogit}. 
This section describes the \pkg{mnlogit} user interface, emphasizing data preparation requirements and model specification via an enhanced formula interface.
To start, we load the package \pkg{mnlogit} in an \proglang{R} session:
\begin{Schunk}
\begin{Sinput}
R> library("mnlogit")
\end{Sinput}
\end{Schunk}

\subsection{Data preparation}
\label{subsection: data preparation}

\pkg{mnlogit} accepts data in the `long' format which requires that if there are $K$ choices, then there be $K$ rows of data for \emph{each} individual (see also Section 1.1 of the \pkg{mlogit} vignette).
Here is a snapshot from data in the `long' format on choice of recreational fishing mode made by 1182 individuals:
\begin{Schunk}
\begin{Sinput}
R> data("Fish", package = 'mnlogit')
R> head(Fish, 8)
\end{Sinput}
\begin{Soutput}
           mode   income     alt   price  catch chid
1.beach   FALSE 7083.332   beach 157.930 0.0678    1
1.boat    FALSE 7083.332    boat 157.930 0.2601    1
1.charter  TRUE 7083.332 charter 182.930 0.5391    1
1.pier    FALSE 7083.332    pier 157.930 0.0503    1
2.beach   FALSE 1250.000   beach  15.114 0.1049    2
2.boat    FALSE 1250.000    boat  10.534 0.1574    2
2.charter  TRUE 1250.000 charter  34.534 0.4671    2
2.pier    FALSE 1250.000    pier  15.114 0.0451    2
\end{Soutput}
\end{Schunk}
In the `Fish' data, there are 4 choices (`beach', `boat', `charter', `pier') available to each individual: labeled by the `chid' (chooser ID).
The `price' and `catch' column show, respectively, the cost of a fishing mode and (in unspecified units) the expected amount of fish caught.
An important point here is that this data varies \emph{both} with individuals and the fishing mode.
The `income' column reflects the income level of an individual and does not vary between choices.
Notice that the snapshot shows this data for two individuals.

The actual choice made by an individual, the `response' variable, is shown in the column `mode'.
\pkg{mnlogit} requires that the data contain a column with exactly two categories whose levels can be coerced to integers by \code{as.numeric()}.
The greater of these integers is automatically taken to mean \code{TRUE}.

The only other column strictly mandated by \pkg{mnlogit} is one listing the names of choices (like column `alt' in Fish data).
However if the data frame is an \code{mlogit.data} class object, then this column maybe \emph{omitted}.
In such cases \pkg{mnlogit} can query the \code{index} attribute of an \code{mlogit.data} object to figure out the information contained in the `alt' column.  

\subsection{Model parametrization}
\label{subsection: parametrization}

Multinomial logit models have a solid basis in the theory of discrete choice models.
The central idea in these discrete models lies in the `utility maximization principle' which states that individuals choose the alternative, from a finite, discrete set, which maximizes a scalar value called `utility'. 
Discrete choice models presume that the utility is completely deterministic for the individual, however modelers can only model a part of the utility (the `observed' part).
Stochasticity entirely arises from the \emph{unobserved} part of the utility. 
Different assumptions about the probability distribution of the unobserved utility give rise to various choice models like multinomial logit, nested logit, multinomial probit, GEV (Generalized Extreme Value), mixed logit etc.
Multinomial logit models, in particular, assume that unobserved utility is i.i.d. and follows a Gumbel distribution.\footnote{See the book~\citet{TRAIN:03}, particularly Chapters 3 and 5, for a full discussion.}

We consider that the \emph{observed} part of the utility for the $i^{th}$ individual choosing the $k^{th}$ alternative is given by: 
\begin{equation}
\label{eqn: unormalized utility}
U_{ik} = \xi_{k} + \dotP{X_i}{\beta_{k}} + \dotP{Y_{ik}}{\gamma_{k}} + \dotP{Z_{ik}}{\alpha}.
\end{equation}
Here Latin letters ($X$, $Y$, $Z$) stand for data while Greek letters ($\xi$, $\alpha$, $\beta$, $\gamma$) stand for parameters.
The parameter $\xi_{k}$ is called the \emph{intercept}.
For many practical applications data in multinomial logit models can be naturally grouped into two types:
\begin{itemize}
\item {\bf Individual specific variables $\vec{X}_{i}$} which does \emph{not} vary between choices (e.g., income of individuals in the `Fish' data of Section~\ref{section: data format and model specification}). 
\item {\bf Alternative specific variables $\vec{Y}_{ij}$ and $\vec{Z}_{ij}$} which vary with alternative and may also differ, for the same alternative, between individuals (e.g., the amount of fish caught in  the `Fish' data: column `catch'). 
\end{itemize}
In \pkg{mnlogit} we model these two data types with three types of coefficients:
\begin{enumerate}
\item Individual specific data with alternative specific coefficients $\dotP{X_{i}}{\beta_{j}}$
\item Alternative specific data with generic coefficients $\dotP{Z_{ik}}{\alpha}$.
\item Alternative specific data with alternative specific coefficients $\dotP{Y_{ik}}{\gamma_{k}}$.
\end{enumerate}
The vector notation serves to remind that more than one variable of each type maybe used to build a model.
For example in the fish data we may choose both the `price' and `catch' with either generic coefficients (the $\vec{\alpha}$) or with alternative specific coefficients (the $\vec{\gamma_{k}}$).

Due to the principle of utility maximization, only differences between utility are meaningful.
This implies that the multinomial logit model can not determine absolute utility.
We must specify the utility for any individual with respect to an arbitrary base value\footnote{In choice model theory this is called `normalizing' the model.} which we choose to be $0$.
For convenience in notation, we fix the choice indexed by $k=0$ as the base, thus \emph{normalized} utility is given by:
\begin{align*}
V_{ik} = U_{ik} - U_{i0} = \xi_{k} - \xi_{0} + \vec{X_i}\cdot(\vec{\beta}_{k} - \vec{\beta}_{0}) + \vec{Y}_{ik}\cdot\vec{\gamma}_{k} - \vec{Y}_{i0}\cdot\vec{\gamma}_{0}+ (\vec{Z}_{ik} - \vec{Z}_{i0})\cdot\vec{\alpha}.
\end{align*}
Notice that the above expression implies that $V_{i0}=0$ $\forall{}i$.
To simplify notation we re-write the normalized utility as:
\begin{align}
\label{eqn: normalized utility}
V_{ik} = \xi_{k} + \vec{X_i}\cdot\vec{\beta}_{k} + \vec{Y}_{ik}\cdot\vec{\gamma}_{k} - \vec{Y}_{i0}\cdot\vec{\gamma}_{0}+ \vec{Z}_{ik}\cdot\vec{\alpha} \qquad k \in [1, K-1]
\end{align}
This equation retains the same \emph{meaning} as the previous, notice the restriction: $k\neq{}0$, since we need $V_{i0} = 0$.
The most significant difference is that $\vec{Z}_{ik}$ in Equation~\ref{eqn: normalized utility}  stands for: $\vec{Z}_{ik} - \vec{Z}_{i0}$ (in terms of  the original data).

The utility maximization principle, together with the assumtion on the error distribution, implies that for multinomial logit models~\citep{TRAIN:03} the probability of individual $i$ choosing alternative $k$, $\Prob_{ik}$, is given by:
\begin{align}
\label{eqn: probability}
\Prob_{ik} = \Prob_{i0}e^{V_{ik}}.
\end{align}
Here $V_{ij}$ is the normalized utility given in Equation~\ref{eqn: normalized utility} and $k=0$ is the base alternative with respect to which we normalize utilities.
The number of available alternatives is taken as $K$ which is a positive integer greater than one.
From the condition that every individual makes a choice, we have that: $\sum_{k=0}^{k=K-1} \Prob_{ik} = 1$,.
This gives us the probability of individual $i$ picking the base alternative:
\begin{align}
\label{eqn: base probability}
\Prob_{i0} = \frac{1}{1+\sum_{k=1}^{K-1}e^{V_{ik}}}.
\end{align}
Note that $K=2$ is the familiar binary logistic regression model.

Equation~\ref{eqn: normalized utility} has implications about which model parameters maybe identified.
In particular for alternative-specific coefficients of individual-specific data we may only estimate the difference $\vec{\beta_{k}} - \vec{\beta_{0}}$.
Similarly for the intercept only the difference $\xi_{k} - \xi_{0}$, and not $\xi_{k}$ and $\xi_{0}$ separately maybe estimated.
For a model with $K$ alternative we estimate $K-1$ sets of parameters $\vec{\beta_{k}} - \vec{\beta_{0}}$ and $K-1$ intercepts $\xi_{k} - \xi_{0}$.


\subsection{Formula interface}
\label{subsection: formula interface}

To specify multinomial logit models in \proglang{R} we need an enhanced version of the standard formula interface - one which is able to handle multi-part formulas.  
In \pkg{mnlogit} we built the formula interface using tools from the \proglang{R} package \pkg{Formula}~\citep{ZEIL:CROIS:10}.
Our formula interface closely confirms to that of the \pkg{mlogit} package.
We illustrate it with examples motivated by the `Fish' dataset (introduced in Section~\ref{section: data format and model specification}).
Consider a multinomial logit model where `price' has a generic coefficient, `income' data being individual-specific has an alternative-specific coefficient and the `catch' also has an alternative-specific coefficient.
That is, we want to fit a model that has the 3 types of coefficients described in Section~\ref{subsection: parametrization}.
Such a model can be specified in \pkg{mnlogit} with a 3-part formula:
\begin{Schunk}
\begin{Sinput}
R> fm <- formula(mode ~ price | income | catch)
\end{Sinput}
\end{Schunk}
By default, the intercept is included, it can be omitted by inserting a `-1' or `0' anywhere in the formula.
The following formulas specify the same model with omitted intercept: 
\begin{Schunk}
\begin{Sinput}
R> fm <- formula(mode ~ price | income - 1 | catch)
R> fm <- formula(mode ~ price | income | catch - 1)
R> fm <- formula(mode ~ 0 + price | income | catch)
\end{Sinput}
\end{Schunk}

We can omit any group of variables from the model by placing a $1$ as a placeholder:
\begin{Schunk}
\begin{Sinput}
R> fm <- formula(mode ~ 1 | income | catch) 
R> fm <- formula(mode ~ price | 1 | catch) 
R> fm <- formula(mode ~ price | income | 1)
R> fm <- formula(mode ~ price | 1 | 1)
R> fm <- formula(mode ~ 1 | 1 | price + catch)
\end{Sinput}
\end{Schunk}
When the meaning is unambiguous, an omitted group of variables need not have a placeholder. 
The following formulas represent the same model where `price' and `catch' are modeled with generic coefficients and the intercept is included:
\begin{Schunk}
\begin{Sinput}
R> fm <- formula(mode ~ price + catch | 1 | 1)
R> fm <- formula(mode ~ price + catch | 1) 
R> fm <- formula(mode ~ price + catch)
\end{Sinput}
\end{Schunk}

\subsection[Using package mnlogit]{Using package \pkg{mnlogit}}
\label{subsection: usage mnlogit}

In an \proglang{R} session with \pkg{mnlogit} loaded, the man page can be accessed in the standard way:
\begin{Schunk}
\begin{Sinput}
R> ?mnlogit
\end{Sinput}
\end{Schunk}
The complete \code{mnlogit} function call looks like:
\begin{Schunk}
\begin{Sinput}
R> mnlogit(formula, data, choiceVar = NULL, maxiter = 50, ftol = 1e-6, gtol
+          = 1e-6, weights = NULL, ncores = 1, na.rm = TRUE, print.level = 0, 
+          linDepTol = 1e-6, start = NULL, alt.subset = NULL, ...)
\end{Sinput}
\end{Schunk}
We have described the `formula' and `data' arguments in previous sections while others are explained in the man page, only the `linDepTol' argument needs further elaboration.
Data used to train the model must satisfy certain necessary conditions so that the Hessian matrix, computed during Newton-Raphson estimation, is full rank (more about this in Appendix~\ref{appendix: data requirements for Hessian non-singularity}).
In \pkg{mnlogit} we use the \proglang{R}  built-in function \code{qr}, with its argument `tol' set to `linDepTol', to check for linear dependencies .
If collinear columns are detected in the data then some are removed so that the remaining columns are linearly independent.

We now illustrate the practical usage of \code{mnlogit} and some of its methods by a simple example.
Consider the model specified by the formula:
\begin{Schunk}
\begin{Sinput}
R> fm <- formula(mode ~ price | income | catch)
\end{Sinput}
\end{Schunk}
This model has:
\begin{itemize}
\item One variable of type $\vec{\alpha}$: `price'.
\item Two variable of type $\vec{\beta_{k}}$: `income' and the intercept.
\item One variable of type $\vec{\gamma_{k}}$: `catch'.
\end{itemize}
In the `Fish' data the number of alternatives  $K=4$, so the number of coefficients in the above model is:
\begin{itemize}
\item $1$ coefficient for data that may vary with individuals and alternatives, corresponding to $\vect{\alpha}$.
\item $2 \times (K-1) = 6$, alternative specific coefficients for individual specific data (note: that we have subtract $1$ from the number of alternative because after normalization the base choice coefficient can't be identified), corresponding to $\vect{\beta_{k}}$.
\item $1 \times K = 4$ alternative specific coefficients for data which may vary with individuals and alternatives, corresponding to $\vect{\gamma_{k}}$.
\end{itemize}
Thus the total number of coefficients in this model is $1 + 6 + 4 = 11$.

We call the function \code{mnlogit} to fit the model using the `Fish' dataset on $2$ processor cores.
\begin{Schunk}
\begin{Sinput}
R> fit <- mnlogit(fm, Fish, ncores=2)
R> class(fit)
\end{Sinput}
\end{Schunk}
\begin{Schunk}
\begin{Soutput}
[1] "mnlogit"
\end{Soutput}
\end{Schunk}
For \code{mnlogit} class objects we have the usual methods associated with \proglang{R} objects: \code{coef}, \code{print}, \code{summary} and \code{predict} methods.
In addition, the returned `fit' object can be queried for details of the estimation process by:
\begin{Schunk}
\begin{Sinput}
R> print(fit$est.stat)
\end{Sinput}
\begin{Soutput}
-------------------------------------------------------------
Maximum likelihood estimation using the Newton-Raphson method
-------------------------------------------------------------
  Number of iterations: 7
  Number of linesearch iterations: 10
At termination: 
  Gradient norm = 2.09e-06
  Diff between last 2 loglik values = 0
  Stopping reason: Succesive loglik difference < ftol (1e-06).
Total estimation time (sec): 0.042
Time for Hessian calculations (sec): 0.005 using 2 processors.
\end{Soutput}
\end{Schunk}
The estimation process terminates when first one of the 3 conditions `maxiter', `ftol' or `gtol' are met.
In case one runs into numerical singularity problems during the Newton iterations, we recommend relaxing `ftol' or `gtol' to obtain a suitable estimate.
The plain Newton method has a tendency to overshoot extrema. 
In \pkg{mnlogit} we have inlcuded a `line search' \footnote{One dimensional minimization along the Newton direction.} which avoids this problem and ensures convergence~\citep{NocedalBook}.

As a convenience, we provide the following method so that an \code{mnlogit} object maybe queried for the number and type of model coefficients.
\begin{Schunk}
\begin{Sinput}
R> print(fit$model.size)
\end{Sinput}
\begin{Soutput}
Number of observations in training data = 1182
Number of alternatives = 4
Intercept turned: ON
Number of parameters in model = 11
  # individual specific variables = 2
  # choice specific coeff variables = 1
  # generic coeff variables = 1
\end{Soutput}
\end{Schunk}

Finally there is provision for hypothesis testing.
We provide the function \code{hmftest} to perform the Hausman-McFadden test for IIA (Independence of Irrelevant Alternatives), which is the central hypothesis underlying multinomial logit models~\citep[Chap. 3]{TRAIN:03}.
Three functions to test for hypotheses, applicable to any model estimated by the maximum likelihood method, are also provided:
\begin{itemize}
\item Function \code{lrtest} to perform the likelihood ratio test.
\item Function \code{waldtest} to perform the Wald test.
\item Function \code{scoretest} to perform the Rao score test.
\end{itemize}
These intent of these tests is succinctly described in Section 6 `Tests' of the \pkg{mlogit} package vignette and we shall not repeat it here.
We encourage the interested user to consult the help page for any of these functions in the usual way, for example the \code{lrtest} help maybe accessed by:
\begin{Schunk}
\begin{Sinput}
R> library("mnlogit")
R> ?lrtest
\end{Sinput}
\end{Schunk}
Functions \code{hmftest} and \code{scoretest} are adapted from code in the \pkg{mlogit} package, while \code{lrtest} and \code{waldtest} are built using tools in the CRAN \proglang{R} package \pkg{lmtest}~\citep{lmtest}. 
 
\section{Estimation algorithm}
\label{section: algorithms and optimization}

In \pkg{mnlogit} we employ maximum likelihood estimation (MLE) to compute model coefficients and use the Newton-Raphson method to solve the optimization problem. 
The Newton-Raphson method is well established for maximizing the logistic family loglikelihoods \citep{HastieTibBook, TRAIN:03}.
However direct approaches of computing the Hessian of multinomial logit model's log-likelihood function have extremely deleterious effects on the computer time and memory required.
We present an alternate approach which exploits structure of the  the intermediate data matrices that arise in Hessian calculation  to achieve the same computation \emph{much} faster while using drastically less memory.
Our approach also allows us to optimally \emph{parallelize} Hessian computation and maximize the use of BLAS (Basic Linear Algebra Subprograms) Level 3 functions, providing an additional factor of speedup.

\subsection{Maximizing the likelihood}
\label{subsection: maximizing the likelihood}

Before going into details we specify our notation.
Throughout we assume that there are $K \geq 3$ alternatives.
The letter $i$ labels individuals (the `choice-makers') while the letter $k$ labels alternatives (the `choices').
We also assume that we have data for $N$ individuals available to fit the model ($N$ is assumed to be much greater than the number of model parameters).
We use symbols in {\bf bold face} to denote matrices, for example $\mat{H}$ stands for the Hessian matrix.

To simplify housekeeping in our calculations we organize model coefficients into a  vector $\vec{\theta}$.
If the intercept is to be estimated then it simply considered another individual specific variable with an alternative specific coefficient but with the special provision that the `data' corresponding to this variable is unity for all alternatives.
The likelihood function is defined by $L(\vect{\theta}) = \prod_{i}\prob{y_i|\vect{\theta}}$, where each $y_i$ labels the  alternative \emph{observed} to chosen by individual $i$.
Now we have:
\[\prob{y_i | \vect{\theta}} = \prod_{k=0}^{K-1} \prob{y_i=k}^{I(y_i=k)}.\]
Here $I(y_i=k)$ is the \emph{indicator function} which unity if its argument is true and zero otherwise.
The likelihood function is given by: $L(\vect{\theta}) = \Pi_{i=1}^{N}L(\vect{\theta}|y_i)$. 
It is more convenient to work with the log-likelihood function which is given by $l(\vect{\theta}) = \text{log} L(\vect{\theta})$.
A little manipulation gives:
\begin{align}
\label{eqn: log-lik function}
 l(\vect{\theta}) = \sum_{i=1}^{N}\left[ -\text{log}\left(1 + \sum_{k=1}^{K-1}\text{exp}(V_{ik})\right) + \sum_{k=1}^{K-1} V_{ik}I(y_i =k)\right].
\end{align}
In the above we make use of the identity $\sum_{k}I(y_i=k) = 1$ and the definition of $\Prob_{i0}$ in Equation~\ref{eqn: base probability}.
\cite{MCFAD:74} has shown that the log-likelihood function given above is globally concave.
A quick argument to demostrate global concavity of $l(\vect{\theta})$ is that it's the sum of affine functions $V_{ik}$ and the negation of the composition of the log-sum-exp function with a set of affine functions.
\footnote{The log-sum-exp function's convexity and its closedness under affine composition are well known, see for example Chapter 3 of \cite{BoydBook}.}

We solve the optimization problem by the Newton-Raphson (NR) method which requires finding a stationary point of the gradient of the log-likelihood.
Note that MLE by the Newton-Raphson method is the same as the Fisher scoring algorithm \citep{HastieTibBook, jialimlogit}.
For our log-likelihood function~\ref{eqn: log-lik function}, this point (which we name $\hat{\theta}$) is unique (because of global concavity) and is given by the solution of the equations:
$\partderiv{l(\vec{\theta})}{\vec{\theta}} = \vec{0}$.
The NR method is iterative and starting at an initial guess obtains an improved estimate of $\hat{\theta}$ by the equation:  
\begin{align}
\label{eqn: NR update}
\vect{\theta}^{new} = \vect{\theta}^{old} - \mat{H}^{-1} \partderiv{l}{\vect{\theta}}.
\end{align}
Here the Hessian matrix, $\mat{H} = \frac{\partial^{2}l}{\partial\vec{\theta}\partial\vec{\theta}^{\prime}}$ and the gradient $\partderiv{l}{\vect{\theta}}$, are both evaluated at $\vect{\theta}^{old}$. 
The vector $\vec{\delta\theta} = -\mat{H}^{-1} \partderiv{l}{\vect{\theta}}$ is called the \emph{full} Newton step.
In each iteration we attempt to update $\vec{\theta}^{old}$ by this amount.
However if the log-likelihood value at the resulting $\vec{\theta}^{new}$ is smaller, then we instead try an update of $\vec{\delta\theta}/2$.
This \emph{linesearch} procedure is repeated with half the previous step until the new log-likelihood value is not lower than the value at  $\vec{\theta}^{old}$.
Using such a linesearch procedure guarantees convergence of the Newton-Raphson iterations ~\citep{NocedalBook}.    

\subsection{Gradient and Hessian calculation}
\label{subsection: gradient and hessian calculation}

Each Newton-Raphson iteration requires computation of the Hessian and gradient of the log-likelihood function.
The expressions for the gradient and Hessian are quite well known\footnote{See for example \citep[Section 2.5]{mlogit} and \citep[Chatper 3]{TRAIN:03}.} and in there usual form are given by: 
\begin{align}
\partderiv{l}{\vect{\theta}} &= \sum_{i}\sum_{k} \left(I(y_i = k) - \Prob_{ik}\right) \tilde{X}_{ik} \nonumber \\
\mat{H} &= -\mat{\tilde{X}}^\top\mat{\tilde{W}}\mat{\tilde{X}}
\label{eqn: old Hessian formula}
\end{align}
For a model where where only individual specific variables are used (that is only the matrix $\mat{X}$ contributes to the utility in Equation~\ref{eqn: normalized utility}), the matrices $\mat{\tilde{X}}$ and $\mat{\tilde{W}}$ are given by \citep{jialimlogit, Bohning:92}:
\begin{align*}
   \mat{\tilde{X}}  & = 
   \begin{pmatrix}
     \mat{X} & \mat{0}       & \mat{0} & \cdots & \mat{0} \\
     \mat{0}       & \mat{X} & \mat{0} & \cdots & \mat{0} \\
     \vdots  & \mat{0}       &   & \ddots & \vdots \\
     \mat{0}       & \cdots  &   & \mat{0}      & \mat{X} \\
    \end{pmatrix},
\end{align*}
here $\mat{X}$ is a matrix of order $N\times{}p$ ($p$ is the number of variables or features) and,
\begin{align*}
   \mat{\tilde{W}}   & = 
   \begin{pmatrix}
     \mat{W_{11}} & \mat{W_{12}} & \cdots & \mat{W_{1,K-1}} \\
    \mat{W_{21}} & \mat{W_{22}} & \cdots & \mat{W_{2,K-1}} \\
     \vdots &   \vdots & \cdots & \vdots   \\
    \mat{W_{K-1,1}} & \cdots & \cdots & \mat{W_{K-1,K-1}} \\
    \end{pmatrix}.
\end{align*}
Here the sub-matrices $\mat{W_{k,t}}$ are \emph{diagonal} matrices of order $N\times{}N$, where $\text{diag}(\mat{W_{k,t}})_{i} = \Prob_{ik}(\delta_{kt} - \Prob_{it})$ and $\delta_{kt}$ is the Kronecker delta which equals $1$ if $k=t$ and $0$ otherwise. 
Using this notation the gradient can be written as~\citep{jialimlogit}:
\begin{align*}
\partderiv{l}{\vec{\theta}} = \mat{\tilde{X}}^\top\left(\vect{y} - \vect{\Prob}\right)
\end{align*}
Where we take vectors $\vect{y}$ and $\vect{\Prob}$ as vectors of length $N \times (K - 1)$, formed by vertically concatenating the $N$ probabilities $\Prob_{ik}$ and responses $I(y_i = k)$,  for each $k\in [1, K-1]$.
The Newton-Raphson iterations of Equation~\ref{eqn: NR update} take the form: 
$\vect{\theta}^{\text{new}}$ = $\vect{\theta}^{\text{old}} + \left(\mat{\tilde{X}}^\top\mat{\tilde{W}}\mat{\tilde{X}}\right)^{-1}\mat{\tilde{X}}(\vect{y} - \vect{\Prob})$.
Although in this section we have shown expressions for models with only individual specific variables, a general formulation of $\mat{\tilde{X}}$ and $\mat{\tilde{W}}$ including the two other types of variables appearing in Equation~\ref{eqn: normalized utility} exists\footnote{And is implemented in the \proglang{R} packages \pkg{mlogit}~\citep{mlogit} and \pkg{VGAM}~\citep{VGAM}.}.
This is presented in Appendix~\ref{appendix: data requirements for Hessian non-singularity} but their specific form is tangential to the larger point we make (our ideas extend to the general case in a simple way). 

An immediate advantage of using the above formulation, is that Newton-Raphson iterations can be carried out using the framework of IRLS (Iteratively Re-weighted Least Squares) \citep[Section 4.4.1]{HastieTibBook}. 
IRLS is essentially a sequence of weighted least squares regresions which offers superior numerical stability compared to explicitly forming $\mat{H}$ and directly solving Equation~\ref{eqn: NR update} \citep[Lecture 19]{TrefethenBook}.
However this method, besides being easy to implement, is computationally very inefficient.
The matrices $\mat{\tilde{X}}$ and $\mat{\tilde{W}}$ are huge, of orders $(K-1)N \times (K-1)p$ and $N(K-1) \times N(K-1)$ respectively, but are otherwise quite sparse and possess a neat structure. 
We now describe our approach of exploiting this structured sparsity.   

\subsection{Exploiting structure - fast Hessian calculation}
\label{subsection: exploiting structure}

We focus our attention on computation of the Hessian since it's the most expensive step, as we later show from empirical measurements in Table~\ref{table: timing profile} of Section~\ref{section: benchmarking performance}.
We start by ordering the vector $\vec{\theta}$, which  is a concatenation of all model coefficients as specified in Equation~\ref{eqn: normalized utility}, in the following manner:
\begin{align}
\label{eqn: theta vec}
\vec{\theta} = \left\lbrace \vec{\beta}_{1}, \vec{\beta}_{2} \dots \vec{\beta}_{K-1}, \vec{\gamma}_{0}, \vec{\gamma}_{1}, \dots  \vec{\gamma}_{K-1}, \vec{\alpha}   \right\rbrace. 
\end{align}
Here, the subscripts index alternatives and the vector notation reminds us there maybe multiple features modeled by coefficients of type $\vec{\beta}$, $\vec{\gamma}$, $\vec{\alpha}$.
In $\vec{\theta}$ we group together coefficients corresponding to an alternative. 
This choice is deliberate and leads to a particular structure of the Hessian matrix of the log-likelihood function with a number of desirable properties.

Differentiating the log-likelihood function with respect to the coefficient vector $\vect{\theta}$, we get:
\begin{align}
\label{eqn: loglik gradient}
\partderiv{l}{\vec{\theta}_m} =
\left\{
  \begin{array}{ll}
  	\mat{M_{m}}^\top\left(\vec{y}_{m} - \vec{\Prob}_{m} \right)  & \mbox{ if } \vec{\theta}_{m} \mbox{ is one of } \left\lbrace \vec{\beta}_{1}, \dots \vec{\beta}_{K-1}, \vec{\gamma}_{0}, \dots  \vec{\gamma}_{K-1} \right\rbrace \\
		\sum_{k=1}\mat{Z_{k}}^\top\left(\vec{y}_{k} - \vec{\Prob}_{k} \right) & \mbox{if } \vec{\theta}_{m} \mbox{ is } \vec{\alpha}
	\end{array}
\right. 
\end{align}
Here we have partitioned the gradient vector into \emph{chunks} according to $\vec{\theta}_{m}$ which is a group of coefficients of a particular type (defined in Section~\ref{subsection: parametrization}), either alternative specific or generic.
Subscript $m$ (and subscript $k$) indicates a particular alternative, for example:
\begin{itemize}
\item if $\vec{\theta}_{m} = \vec{\beta}_{1}$ then $m=1$
\item if $\vec{\theta}_{m} = \vec{\beta}_{K-1}$ then $m=K-1$
\item if $\vec{\theta}_{m} = \vec{\gamma}_{1}$ then $m=1$.
\end{itemize}
The vector $\vec{y}_{m}$ is a vector of length $N$ whose $i^{th}$ entry is given by $I(y_{i} = m)$, it tells us whether the observed choice of individual $i$ is alternative $m$, or not.
Similarly $\vec{\Prob}_{m}$ is vector of length whose $i^{th}$ entry is given by $\Prob_{im}$, which is the probability individual $i$ choosing alternative $m$.
The matrices $\mat{M_{m}}$ and $\mat{Z_{k}}$ contain data for choice $m$ and $k$, respectively. 
Each of these matrices has $N$ rows, one for each individual.
Specifically: 
\begin{align*}
\mat{M_{m}} &= \mat{X} \qquad \mbox{if } \vec{\theta}_{m} \in \left\lbrace \vec{\beta}_{1}, \dots \vec{\beta}_{K-1}\right\rbrace \\
\mat{M_{m}} &= \mat{Y_{m}} \qquad \mbox{if }\vec{\theta}_{m} \in \left\lbrace \vec{\gamma}_{0}, \dots  \vec{\gamma}_{K-1} \right\rbrace.
\end{align*}
Similarly, the matrices $\mat{Z}_{k}$ are analogues of the $\mat{Y}_{m}$  and have $N$ rows each (note that due to normalization $\mat{Z_{0}} = \mat{0}$).

To compute the Hessian we continue to take derivatives with respect to chunks of coefficients $\vec{\theta}_{m}$.
On doing this we can write the Hessian in a very simple and compact \emph{block} format as shown below.
\begin{align}
\label{eqn: hessian block}
\mat{H_{nm}}=\frac{\partial^{2}l}{\partial\vec{\theta}_{n}\partial\vec{\theta}_m^{\prime}} =
\left\{
	\begin{array}{lll}
		-\mat{M_{n}}^\top\mat{W_{nm}}\mat{M_{m}}  & \mbox{ if } \vec{\theta}_{n}, \vec{\theta_{m}} \in \left\lbrace \vec{\beta}_{1}, \dots \vec{\beta}_{K-1}, \vec{\gamma}_{0}, \dots  \vec{\gamma}_{K-1} \right\rbrace \\
		-\sum_{k=1}\mat{M_{n}}^\top\mat{W_{nk}}\mat{Z_{k}} & \mbox{if } \vec{\theta}_{n} \in  \left\lbrace \vec{\beta}_{1}, \dots  \vec{\gamma}_{K-1} \right\rbrace \mbox{ \& } \vec{\theta}_{m} \mbox{ is } \vec{\alpha} \\
		-\sum_{k,t=1} \mat{Z_{k}}^\top\mat{W_{kt}}\mat{Z_{t}} & \mbox{if } \vec{\theta}_{n} \mbox{ is }  \vec{\alpha}  \mbox{ \& } \vec{\theta}_{m} \mbox{ is } \vec{\alpha}
	\end{array}
\right. 
\end{align}
Here $\mat{H_{nm}}$ is a block of the Hessian and the matrices $\mat{W_{nm}}$ are \emph{diagonal} matrix of dimension $N\times{}N$, whose $i^{th}$ diagonal entry is given by: $\Prob_{in}(\delta_{nm} - \Prob_{im})$.\footnote{Here $\delta_{nm}$ is the Kronecker delta, which is 1 if $n=m$ and 0 otherwise.}
The details of taking derivatives in this block-wise fashion are given in Appendix~\ref{appendix: log-likelihood differentiation }.

The first thing to observe about Equation~\ref{eqn: hessian block} is that effectively utilizes spartsity in the matrices $\mat{\tilde{X}}$ and $\mat{\tilde{W}}$ to obtain very \emph{compact} expressions for $\mat{H}$.
The block format of the Hessian matrix is particularly suited for extremely efficient numerical computations.
Notice that each block can be computed \emph{independently} of other blocks with two matrix multiplications.
The first of these involves multiplying a diagonal matrix to a dense matrix, while the second requires multiplication of two dense matrices.
We handle the first multiplication with a handwritten loop which exploits the sparsity of the diagonal matrix, while the second multiplication is handed off to a BLAS (Basic Linear Algebra Subprograms) call for optimal efficiency~\citep{GolubBook}\footnote{Hessian computation is implemented in a set of  optimized \proglang{C++} routines}.
Computing the Hessian block-by-block is very efficient since we can use level 3 BLAS calls (specifically \code{DGEMM}) to handle the most intensive calculations.
Another useful property of the Hessian blocks is that because matrices $\mat{W_{nm}}$ are diagonal (hence symmetric), we have the \emph{symmetry property} $\mat{H_{nm}} = \mat{H_{mn}}^\top$, implying that we only need to compute roughly \emph{half} of the blocks.

Independence of Hessian blocks leads to a very fruitul strategy for \emph{parallelizing} Hessian calculations: we simply divide the work of computing blocks in the upper triangular part of the Hessian among available threads.
This strategy has the great advantage that threads don't require any synchronization or communication overhead. 
However the cost of computing all Hessian blocks is not the same: the blocks involving generic coefficients (the $\vec{\alpha}$) take much longer to compute longer.
In \pkg{mnlogit} implementation, computation of the blocks involving generic coefficients is handled separately from other blocks and is optimized for serial computation. 

Hessian calculation is, by far, the most time consuming step in solving the multinomial logit MLE problem via the Newton-Raphson method.
The choice we make in representing the Hessian in the block format of Equation~\ref{eqn: hessian block} has dramatic effects on the \emph{time} (and memory) it takes for model estimation.
In the next section we demonstrate the impact on computation times of this choice when contrasted with earlier approaches.


\section{Benchmarking performance}
\label{section: benchmarking performance}

For the purpose of performance profiling \pkg{mnlogit} code, we use simulated data generated using a custom R function \code{makeModel} sourced from \code{simChoiceModel.R} which is available in the folder \code{mnlogit/vignettes/}.
Using simulated data we can easily vary problem size to study performance of the code - which is our main intention here - and make comparisons to other packages.
Our tests have been performed on a dual-socket, 64-bit Intel machine with 8 cores per socket which are clocked at $2.6$ GHz\footnote{The machine has 128 GB of RAM and 20 MB of shared L3 cache.}.
\proglang{R} has been natively compiled on this machine using \code{gcc} with BLAS/LAPACK support from single-threaded Intel MKL v11.0.

The 3 types of model coefficients mentioned in Section~\ref{subsection: parametrization} entail very different computational requirements.
In particular it can be seen from Equations~\ref{eqn: loglik gradient} and~\ref{eqn: hessian block}, that Hessian and gradient calculation is computationally  very demanding  for generic coefficients.
For clear-cut comparisons we speed test the code with 4 types of problems described below. 
In our results we shall use $K$ to denote the number of alternatives and $n_{p}$ to denote the total number of coefficients in the model.
\begin{enumerate}
\item {\bf Problem `X':} A model with only individual specific data with alternative specific coefficients.
\item {\bf Problem `Y':} A model with data varying both with individuals and alternatives and alternative specific model coefficients.
\item {\bf Problem `Z':} Same type of data as problem `Y' but with generic coefficients which are independent of alternatives.
\item {\bf Problem `YZ':} Same type of data as problem `Y' but with a mixture of alternative specific and generic coefficients.
\end{enumerate}
Although problem `X' maybe considered a special case of problem `Y' but we have considered it separately, because it's typically used in machine learning domains as the simplest linear multiclass classifier~\citep{HastieTibBook}. 
We shall also demonstrate that \pkg{mnlogit} runs much faster for this class of problems.\footnote{And use it to compare with the \proglang{R} packages \pkg{VGAM} and \pkg{nnet} which run \emph{only} this class of problems (see Appendix~\ref{appendix: timing tests}).}
The `YZ' class of problems serves to illustrate a common use case of multinomial logit models in econometrics where the data may vary with both individuals and alternatives while the coefficients are a mixture of alternative specific and generic types (usually only a small fraction of variables are modeled with generic coefficients). 

The workings of \pkg{mnlogit} can be logically broken up into 3 steps:
\begin{enumerate}
\item Pre-processing: Where the model formula is parsed and matrices are assembled from a user supplied \code{data.frame}. 
We also check the data for collinear columns (and remove them) to satisfy certain necessary conditions, specified in Appendix~\ref{appendix: data requirements for Hessian non-singularity}, for the Hessian to be non-singular.
\item Newton-Raphson Optimization: Where we maximize the log-likelihood function to estimate model coefficients.
This  involves solving a linear system of equations and one needs to compute the log-likelihood function's gradient vector and  Hessian matrix.
\item Post-processing: All work needed to take the estimated coefficients and returning an object of class \code{mnlogit}.
\end{enumerate}
Table~\ref{table: timing profile} has profile of \pkg{mnlogit} performance for the four representative problems discussed earlier. 
Profiling the code clearly shows the highest proportion of time is spent in Hessian calculation (except for problem `Z', for which the overall time is relatively lower).
This is not an unexpected observation, it underpins our focus on optimizing Hessian calculation.
\begin{table}[t]
\centering
\begin{tabular}{|c|c|c|c|c|c|}
\hline
Problem & Pre-processing time(s) & NR time(s) & Hessian time(s)  & Total time(s) & $n_{p}$                              \\ \hline
X  & 93.64  & 1125.5 & 1074.1 & 1226.7 & 4950  \\ \hline  	
Y  &  137.0 & 1361.5 & 1122.4 & 1511.8 & 5000  \\ \hline	
Z  &  169.9 & 92.59  & 60.05  & 272.83 & 50    \\ \hline
YZ &	 170.1  & 1247.4 & 1053.1 & 1417.5 & 4505  \\ \hline
\end{tabular}
\caption{ 
Performance profile of \pkg{mnlogit} for different problems with $50$ variables and $K=100$ alternatives with data for $N=100,000$ individuals. 
All times are in seconds.
`NR time' is the total time taken in Newton-Raphson estimation while `Hessian time' (which is included in `NR time') is the time spent in computing Hessian matrices.
Column $n_{p}$ has the number of model coefficients.
Problem `YZ' has $45$ variables modeled with individual specific coefficients while the other $5$ variables are modeled with generic coefficients.
}
\label{table: timing profile}
\end{table}
Notice the very high pre-processing time for problem `Z' whereof a large portion is spent in ensuring that the data satisfies necessary conditions, mentioned in Appendix~\ref{appendix: data requirements for Hessian non-singularity}, for the Hessian to be non-singular.

\subsection[Comparing mnlogit performance]{Comparing \pkg{mnlogit} performance}
\label{subsection: comparing mnlogit performance}

We now compare the performance of \pkg{mnlogit} in \emph{single-threaded} mode with some other \proglang{R} packages.
This section focuses on the comparison with the \proglang{R} package \pkg{mlogit} since it's the only one which covers the entire range of variable and data types as \pkg{mnlogit}.
Appendix~\ref{appendix: timing tests} contains a synopsis of our data generation and timing methods including a comparison of \pkg{mnlogit} with the \proglang{R} packages \pkg{VGAM} and \pkg{nnet}.

Table~\ref{table: relative times with mlogit} shows the ratio between \pkg{mlogit} and \pkg{mnlogit} running times for the 4 classes of problems considered in Table~\ref{table: timing profile}.
We see that for most problems, except those of type `Z', \pkg{mnlogit} outperforms \pkg{mlogit} by a large factor.
\begin{table}[t!]
\centering
\begin{tabular}{|c||c|c|c||c|c|c|}
\hline
\textbf{Optimizer} & \multicolumn{3}{|c||}{\textbf{Newton-Raphson}} & \multicolumn{3}{|c|}{\textbf{BFGS}} \\ \hline
\textbf{K}       & \textbf{10} & \textbf{20} & \textbf{30} & \textbf{10} & \textbf{20} & \textbf{30} \\ \hline
Problem X  & 18.9 &  37.3 & 48.4 &  14.7 &  29.2 & 35.4  \\ \hline  	
Problem Y  &	13.8 &	20.6	& 33.3  &  14.9 &  18.0 & 23.9	\\ \hline	
Problem YZ & 10.5 &	22.8	& 29.4	 &  10.5 &  17.0 & 20.4   \\ \hline
Problem Z & 1.16 &	1.31	& 1.41	 &  1.01 &  0.98 & 1.06   \\ \hline
\end{tabular}
\caption{ 
Ratio between \pkg{mlogit} and \pkg{mnlogit} total running times on a single processor for problems of various sizes and types. 
Each problem has $50$ variables with $K$ alternatives and $N = 50*K*20$ observations to train the model.
\pkg{mlogit} has been run separately with two optimizers: Newton-Raphson and BFGS.
In all cases the iterations terminated when the difference between log-likelihoods in successive iterations reduced below $10^{-6}$.
Note: These numbers can vary depending on the BLAS implementation linked to \proglang{R} and hardware specifications. 
}
\label{table: relative times with mlogit}
\end{table}
We have not run larger problems for this comparison because \pkg{mlogit} running times become too long, except problem `Z'\footnote{In this case with $K=100$ and keeping other parameters the same as Table~\ref{table: relative times with mlogit}, \pkg{mnlogit} outperforms \pkg{mlogit} by factors of $1.35$ and $1.26$ while running the NR and BFGS, respectively.}.

Besides Newton-Raphson, which is the default, we have also run \pkg{mlogit} with the BFGS optimizer.
Typically the BFGS method, part of the quasi-Newton class of methods, takes more iterations than the Newton method but with significantly lower cost per iteration since it never directly computes the Hessian matrix.
Typically for large problems the cost of computing the Hessian becomes too high and the BFGS method  becomes overall faster than the Newton method~\citep{NocedalBook}.
Our approach in \pkg{mnlogit} attacks this weakness of the Newton method by exploiting the structure and sparsity in matrices involved in the Hessian calculation to enable it to outperform BFGS.

\subsection{Parallel performance}
\label{subsection: Parallel performance}

\begin{figure}[t]
\centering \includegraphics[width=0.8\textwidth]{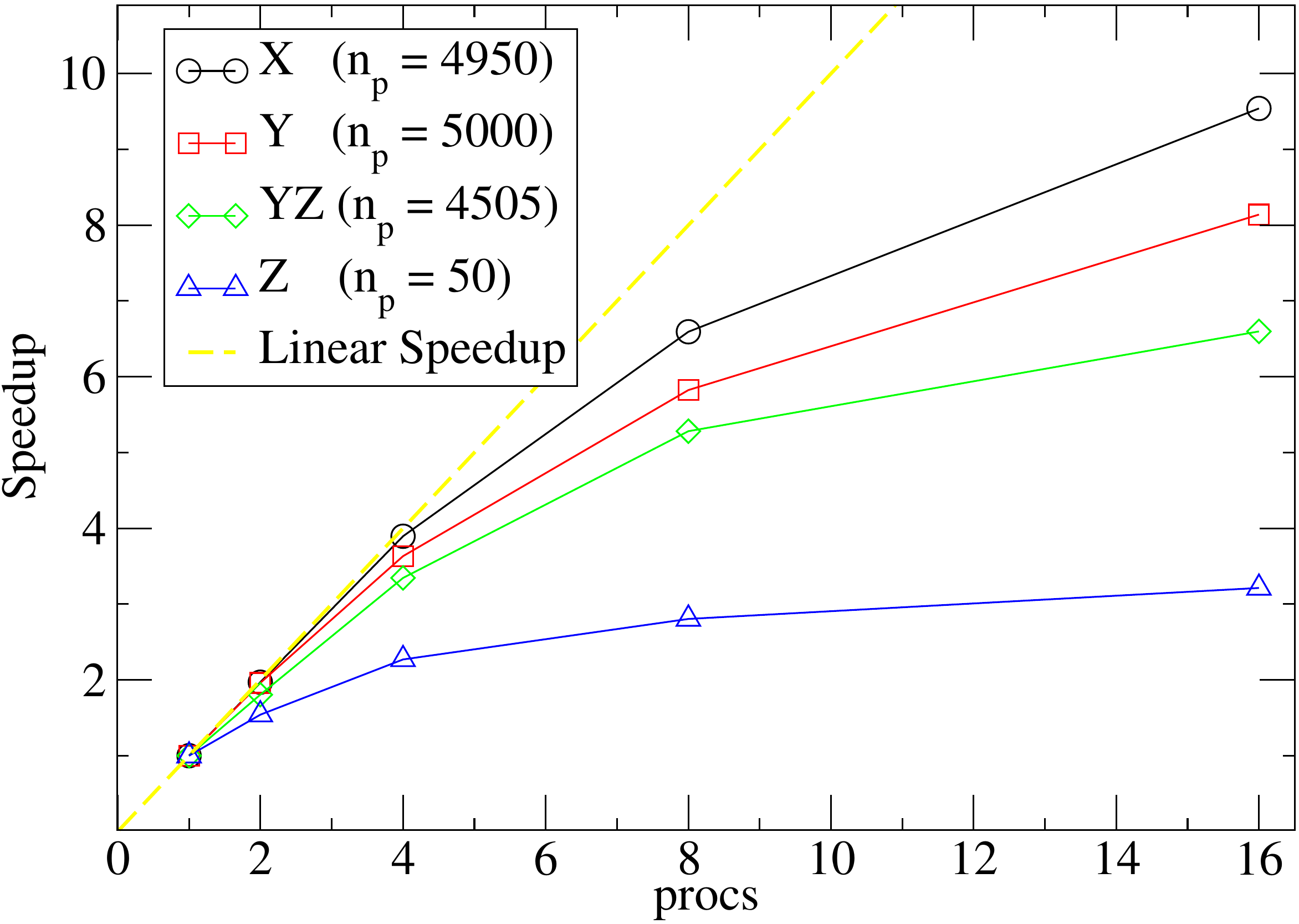}
\caption{
Parallel Hessian calculation speedup (ratio of parallel to single thread running time) for 2, 4, 8, 16, processor cores for problems of Table~\ref{table: timing profile}.
The dashed `Linear Speedup' guideline represents perfect parallelization.
}
\label{fig: hessian parallel speedup}
\end{figure}

We now now turn to benchmarking \pkg{mnlogit's} parallel performance. 
Figure~\ref{fig: hessian parallel speedup} shows the speedups we obtain in Hessian calculation for the same problems considered in Table~\ref{table: timing profile}.
The value of $n_p$, the number of model parameters, is significant because it's the dimension of the Hessian matrix  (the time taken to compute the Hessian scales like $\text{O}(n_{p}^{2})$).
We run the parallel code separately on 2, 4, 8, 16 processor cores, comparing in each case with the single core time. 
Figure~\ref{fig: hessian parallel speedup} shows that it's quite profitable to parallelize problems `X' and `Y', but the gains for problem 'Z' are not too high.
This is because of a design choice we make: Hessian calculation for type `Z' variables is optimized for serial processing. 
In practical modeling, number of model coefficients associated with `Z' types variable is not high, especially when compared to those of types `X' and `Y' (because the number of the coefficients of these types is also proportional to the number of choices in the model, see Section~\ref{subsection: usage mnlogit}).
For problems of type `YZ' (or other combinations which involve `Z'), parallelization can bring significant gains if the number of model coefficients of type `Z' is low, relative to other types.

It can also be seen from figure~\ref{fig: hessian parallel speedup} that, somewhat mysteriously,  parallel performance degrades quickly as the number of processor cores is increased beyond $4$. 
This is a consequence of the fact that our OpenMP progam runs on a machine with shared cache and main memory, so increasing the number of threads degrades performance by increasing the chance of cache misses and hence slowing memory lookups.
This is an intrinsic limitation of our hardware for which there is a  \emph{theoretically simple} solution: run the program on a machine with a larger cache!

An important factor to consider in parallel speedups of the \emph{whole program} is Amdahl's Law\footnote{See Chapter 6 of~\cite{openmpBook}.} which limits the maximum speedup that maybe be achieved by any parallel program.
Assuming parallelization between $n$ threads, Amdahl's law states that the ultimate speedup is given by: $S_{n} = \frac{1}{f_{s} + \left(1 - f_{s}\right)/n}$, where $f_s$ is the fraction of non-parallelized, serial code.
Table~\ref{table: S infinty} lists the observed speedups on $2$, $4$ and $8$ processor coress  together with $f_{s}$ for problems of Table~\ref{table: timing profile}.
\begin{table}[h]
\centering
\begin{tabular}{|c|c|c|c|c|}
\hline
Problem & Serial fraction ($f_s$) & $S_{2}$ & $S_{4}$ & $S_{8}$  \\ \hline
X  & 0.124 & 1.76(1.78) & 2.87(2.92) & 4.04(4.28)                \\ \hline  	
Y  & 0.258 & 1.59(1.62) & 2.26(2.27) & 2.59(2.85)                \\ \hline	
Z  & 0.780 & 1.08(1.12) & 1.14(1.20) & 1.17(1.24)  	            \\ \hline
YZ & 0.257 & 1.44(1.59) & 2.08(2.26) & 2.36(2.86)                \\ \hline
\end{tabular}
\caption{ 
Parallel speedup of \pkg{mnlogit} versus serial performance, (parentheses: predicted ultimate speedup from Amdahl's law) for problems of table~\ref{table: timing profile}.  
$S_{2}$, $S_{4}$ and $S_{16}$ are observed speedups on $2$, $4$ and $16$ processor cores respectively, while $f_{s}$ is the estimated fraction of time spent in the serial portion of the code. 
}
\label{table: S infinty}
\end{table}
We take the time \emph{not} spent in computing the Hessian as the `serial time' to compute $f_{s}$ and neglect the serial time spent in setting up the parallel computation in Hessian calculation, which mainly involves spawning threads in OpenMP and allocating separate blocks of working memory for each thread\footnote{For type `Z' problems, this is an underestimate because some Hessian calculation is also serial.}.
Our tests have shown that compared to the Hessian calculation, the (serial) work required in setting up parallel computation is negligible, except for very small problems.


\section{Discussion}
\label{section: discussion}

Through \pkg{mnlogit} we seek to provide the community a package which combines quick calculation and the ability to handle data collinearity with a software interface which encompasses a wide range of multinomial logit models and data types used in econometrics and machine learning.
Our main idea, exploiting matrix structure in large scale linear algebra calculations is not novel; however this work is the first, as far as we are aware, to apply it to the estimation of multinomial logit models  problems in a working software package.
The parallelization capability of \pkg{mnlogit}, which can easily add a 2x-4x factor of speedup on now ubiquitous mulitcore computers, is another angle which is underutilized in statistical software.
Although \pkg{mnlogit} code is not parallelized to the maximum possible extent, parallelizing the most expensive parts of the calculation was an important design goal. 
We hope that practical users of the package benefit from this feature.

This work was initially motivated by the need to train large-scale multinomial logistic regression models.
For very large-scale problems, Newton's method is usually outperformed by gradient based, quasi-Newton methods like the l-BFGS algorithm~\citep{l-BFGS1989}.
Hessian based methods based still hold promise for such problems.
The class of inexact Newton (also called truncated Newton) methods are specifically designed for problems where the Hessian is expensive to compute but taking a Hessian-vector product (for any given vector) is much cheaper~\citep{NashSurvey2000}.
Multinomial logit models have a Hessian with a structure which permits taking cheap, implicit products with vectors.
Where applicable, inexact Newton methods have the promise of being better than l-BFGS methods~\citep{NasNocedal1991} besides having low memory requirements (since they never store the Hessian) and are thus very scalable.
In the future we shall apply inexact Newton methods to estimating multinomial logit models to study their convergence properties and performance.

\section*{Acknowledgements}

We would like to thank Florian Oswald of the Economics Department at University College London for contributing code for the \code{predict.mnlogit} method.
We are also grateful to numerous users of \pkg{mnlogit} who gave us suggestions for improving the software and reported bugs.

\appendix

\section{Log-likelihood differentiation}
\label{appendix: log-likelihood differentiation }

In this Appendix we give the details of our computation of gradient and Hessian of the log-likelihood function in Equation~\ref{eqn: log-lik function}.
We make use of the notation of Section~\ref{subsection: exploiting structure}. 
Taking the derivative of the log-likelihood with respect to a \emph{chunk} of coefficient $\vec{\theta}_{m}$ one gets:
\begin{align*}
\partderiv{l}{\vec{\theta}_m} = \sum_{i=1}^{N} \left[ \frac{1}{\Prob_{i0}} \partderiv{\Prob_{i0}}{\vec{\theta}_{m}} + \sum_{k=1}^{K-1}I(y_i = k)\partderiv{V_{ik}}{\vec{\theta}_{m}} \right].
\end{align*}
The second term in this equation is a constant term, since the utility $V_{ik}$, defined in Equation~\ref{eqn: normalized utility}, is a linear function of the coefficients. 
Indeed we have:
\begin{align}
\label{eqn: gradient of utility}
 \sum_{i=1}^{N}\sum_{k=1}^{K-1}I(y_i = k) \partderiv{V_{ik}}{\vect{\theta}_{m}} =
\left\{
	\begin{array}{ll}
		\mat{M_{m}}^\top\vect{y}_{m}& \mbox{ if } \vec{\theta}_{m} \in \left\lbrace \vec{\beta}_{1}, \dots \vec{\beta}_{K-1}, \vec{\gamma}_{0}, \dots  \vec{\gamma}_{K-1} \right\rbrace \\
		 \sum_{k=1}\mat{Z_{k}}^\top\vect{y_{k}} & \mbox{if } \vec{\theta}_{m} \mbox{ is } \vec{\alpha}
	\end{array}
\right. 
\end{align}
The vectors $\vect{y}_{m}$ and the matrices $\mat{M_{m}}$ and $\mat{Z_{k}}$ are specified in Section~\ref{subsection: exploiting structure}.
We take the derivative of the base case probability, which is specified in Equation~\ref{eqn: base probability}, as follows:
\begin{align}
\label{eqn: base prob deriv}
\sum_{i=1}^{N}\frac{1}{\Prob_{i0}}\partderiv{\Prob_{i0}}{\vec{\theta}_{m}} =
\left\{
	\begin{array}{ll}
		-\mat{M_{m}}^\top\cdot\vec{\Prob}_{m}& \mbox{ if } \vec{\theta}_{m} \in \left\lbrace \vec{\beta}_{1}, \dots \vec{\beta}_{K-1}, \vec{\gamma}_{0}, \dots  \vec{\gamma}_{K-1} \right\rbrace \\
		 -\sum_{k=1}\mat{Z_{k}}^\top\vec{\Prob}_{k} & \mbox{if } \vec{\theta}_{m} \mbox{ is } \vec{\alpha}
	\end{array}
\right. 
\end{align}
Here the probability vector $\vect{\Prob}_{m}$ is of length $N$ with entries $\Prob_{im}$.
In the last line we have used the fact that, after normalization, $\mat{Z}_{0}$ is $\mat{0}$.
Using Equations~\ref{eqn: gradient of utility} and~\ref{eqn: base prob deriv} we get the gradient in the form shown in Equation~\ref{eqn: loglik gradient}.

Upon differentiating the probability vector $\vec{\Prob}_{k}$ ($k\geq 1$) in Equation~\ref{eqn: probability} with respect to $\vec{\theta}_{m}$ we get:
\begin{align}
\label{eqn: probability deriv}
\partderiv{\vec{\Prob}_{k}}{\vec{\theta}_{m}} = 
\left\{
  \begin{array}{ll}
		\mat{W_{km}}\mat{M_{m}}  & \mbox{ if } \vec{\theta}_{m} \in \left\lbrace \vec{\beta}_{1}, \dots \vec{\beta}_{K-1}, \vec{\gamma}_{0}, \dots  \vec{\gamma}_{K-1} \right\rbrace \\
		\mat{D}(\vec{\Prob}_{k}) \left( \mat{Z_{k}} - \sum_{t=1}\mat{Z_{t}}\mat{D}(\vec{\Prob}_{t})\right) & \mbox{if } \vec{\theta}_{m} \mbox{ is } \vec{\alpha}
	\end{array}
\right. 
\end{align}
where $\mat{D}(\vec{\Prob}_{k})$ is an $N\times{}N$ \emph{diagonal matrix}
whose $i^{th}$ diagonal entry is $\Prob_{ik}$ and, matrix $\mat{W_{km}}$ is also an an $N\times{}N$ \emph{diagonal matrix} whose $i^{th}$ diagonal entry is $\Prob_{ik}(\delta_{km} - \Prob_{im})$. 
In matrix form this is: $\mat{W_{km}} =  \delta_{km}\mat{D}(\vec{\Prob}_{k})  - \mat{D}(\vec{\Prob}_{k})\mat{D}(\vec{\Prob}_{m})$ where $\delta_{km}$ is the Kronecker delta.

We write the Hessian of the log-likelihood in \emph{block} form as:
\begin{align*}
\mat{H_{nm}}=\frac{\partial^{2}l}{\partial\vec{\theta}_{n}\partial\vec{\theta}_{m}^{\prime}} = \sum_{i=1}^{N} \left[\frac{1}{\Prob_{i0}}\frac{\partial^2 \Prob_{i0}}{\partial\vec{\theta}_{n}\partial\vec{\theta}_{m}^{\prime}} - \frac{1}{\Prob_{i0}^{2}} \partderiv{\Prob_{i0}}{\vec{\theta}_{n}}  \partderiv{\Prob_{i0}}{\vec{\theta}_{m}}\right].
\end{align*}
However it can be derived in a simpler way by differentiating the gradient with respect to $\vec{\theta}_{n}$.
Doing this and making use of Equation~\ref{eqn: probability deriv} gives us Equation~\ref{eqn: hessian block}.
The first two cases of equation are fairly straightforward with the matrices $\mat{W}_{km}$ being the same as shown in Equation~\ref{eqn: probability deriv}.
The third case, when ($\vec{\theta}_{n}, \vec{\theta}_{m}$ are both $\vec{\alpha})$,  is a bit messy and we describe it here.
\begin{align*}
\mat{H_{nm}} &= -\sum_{k=1}^{K-1}\left[ \mat{Z_{k}}^\top\mat{D}(\vec{\Prob}_{k}) \left( \mat{Z_{k}} - \sum_{t=1}^{K-1}\mat{D}(\vec{\Prob}_{t})\mat{Z_{t}} \right) \right] \\
&= -\sum_{k=1}^{K-1}\sum_{t=1}^{K-1} \mat{Z_{k}}^\top \left[ \delta_{kt}\mat{D}(\vec{\Prob}_{k})  - \mat{D}(\vec{\Prob}_{k})\mat{D}(\vec{\Prob}_{t})  \right] \mat{Z_{t}} \\
&= -\sum_{k=1}\sum_{t=1} \mat{Z_{k}}^\top \mat{W_{kt}} \mat{Z_{t}}.
\end{align*}
Here the last line follows from the definition of matrix $\mat{W_{kt}}$ as used in Equation~\ref{eqn: probability deriv}.

\section{Data requirements for Hessian non-singularity}
\label{appendix: data requirements for Hessian non-singularity}

We derive necessary conditions on the data for the Hessian to be non-singular.
Using notation from Section~\ref{subsection: gradient and hessian calculation}, we start by building a `design matrix' $\mat{\tilde{X}}$ by concatenating data matrices $\mat{X}$, $\mat{Y_{k}}$ and  $\mat{Z_{k}}$ in the following format:
\begin{align}
\label{eqn: IRLS design mat}
   \mat{\tilde{X}}  & = 
   \begin{pmatrix}
     \mat{X} & 0 & \cdots & 0 & 0 & 0 & \cdots & 0 & \mat{Z_{1}}/2\\
     0 & \mat{X} &  \cdots & 0 & 0 & 0 & \cdots & 0 & \mat{Z_{2}}/2\\
     \vdots &  &   \ddots & \vdots & \vdots & \vdots & \vdots & \vdots & \vdots   \\
     0 & \cdots & 0 & \mat{X} & 0 & 0 & \cdots & 0 & \mat{Z_{K-1}}/2 \\
     0 & \cdots & \cdots & 0 & \mat{Y_{0}} & 0 & \cdots & 0 & 0 \\
     0 & \cdots & \cdots & 0 & 0 & \mat{Y_{1}} & \cdots & 0 & \mat{Z_{1}}/2\\
     0 & \cdots & \cdots & 0 & 0 & 0 & \ddots & 0 & \vdots \\
     0 & \cdots & \cdots & 0 & 0 & 0 & \cdots & \mat{Y_{K-1}} & \mat{Z_{K-1}}/2 \\
    \end{pmatrix}.
\end{align}
In the above $0$ stands for a matrix of zeros of appropriate dimension.
Similarly we build two more matrices $\mat{Q}$ and $\mat{Q_{0}}$ as shown below:
\begin{align*}
   \mat{Q}   & = 
   \begin{pmatrix}
     \mat{W_{11}} & \mat{W_{12}} & \cdots & \mat{W_{1,K-1}} \\
    \mat{W_{21}} & \mat{W_{22}} & \cdots & \mat{W_{2,K-1}} \\
     \vdots &   \vdots & \cdots & \vdots   \\
    \mat{W_{K-1,1}} & \cdots & \cdots & \mat{W_{K-1,K-1}} \\
    \end{pmatrix},
\end{align*}
\begin{align*}
   \mat{Q_{0}}   & = 
   \begin{pmatrix}
     \mat{W_{10}}  \\
    \mat{W_{20}}  \\
     \vdots    \\
    \mat{W_{K-1,0}}  \\
    \end{pmatrix}.
\end{align*}
Using the 2 matrices above we define a `weight' matrix $\mat{\tilde{W}}$:
\begin{align}
\label{eqn: IRLS weight mat}
   \mat{\tilde{W}}  & = 
   \begin{pmatrix}
     \mat{Q} & \mat{Q_{0}} & \mat{Q} \\
     \mat{Q_{0}}^\top & \mat{W_{00}} & \mat{Q_{0}}^\top \\
     \mat{Q} & \mat{Q_{0}} & \mat{Q} \\
    \end{pmatrix},  
\end{align}
The full Hessian matrix, containing all the blocks of Equation~\ref{eqn: hessian block}, is given by: $\mat{H} = \mat{\tilde{X}}^\top \mat{\tilde{W}}\mat{\tilde{X}}$.
For the matrix $\mat{H}$ to be non-singular, we must have the matrix $\mat{\tilde{X}}$ be full-rank.
This leads us to the following \emph{necessary conditions} on the data for the Hessian to be non-singular:
\begin{enumerate}
\item All matrices in the set: $\{\mat{X}$, $\mat{Y_{0}}$, $\mat{Y_{1}}$ $\dots$ $\mat{Y_{K-1}}\}$ must be of full rank.
\item Atleast one matrix from the set: $\{ \mat{Z_{1}}, \mat{Z_{2}}$  \text{\dots} $\mat{Z_{K-1}}\}$ must be of full rank.
\end{enumerate}
In \pkg{mnlogit} we directly test condition 1, while the second condition is tested by checking for collinearity among the columns of the matrix\footnote{Since number of rows is less than the number of columns}:
   \[\begin{pmatrix}
     \mat{Z_{1}}\quad \mat{Z_{2}}\quad \hdots\quad \mat{Z_{K-1}}
    \end{pmatrix}^\top.\]
Columns are arbitrarily dropped one-by-one from a collinear set until the remainder becomes linearly independent.

{\bf Another necessary condition:} It can be shown with some linear algebra manipulations (omitted because they aren't illuminating) that if we have a model with has \emph{only}: data for generic variables independent of alternatives \emph{and} the intercept, then the resulting Hessian will always be singular. 
\pkg{mnlogit} does not attempt to check the data for this condition which is independent of the 2 necessary conditions given above.

\section{Timing tests}
\label{appendix: timing tests}

We give the details of our simulated data generation process and how we setup runs of the \proglang{R} packages \pkg{mlogit}, \pkg{VGAM} and \pkg{nnet} to compare running times against \pkg{mnlogit}.
We start by loading \pkg{mlogit}  into an \proglang{R} session:
\begin{Schunk}
\begin{Sinput}
R> library("mlogit")
\end{Sinput}
\end{Schunk}
Next we generate data in the `long format' (described in Section~\ref{section: data format and model specification}) using the \code{makeModel} function sourced from the file \code{simChoiceModel.R} which is in the \code{mnlogit/vignettes/} folder.
The data we use for the timing tests shown here is individual specific (problem `X' of Section~\ref{section: benchmarking performance}) because this is the only one that packages \pkg{VGAM} and \pkg{nnet} can run.
We generate data for a model with $K$ choices as shown below.
\begin{Schunk}
\begin{Sinput}
R> source("simChoiceModel.R")
R> data <- makeModel('X', K=10)
\end{Sinput}
\end{Schunk}
Default arguments of \code{makeModel} set the number of variables and the number of observations, which are:
\begin{Schunk}
\begin{Soutput}
Number of choices in simulated data = K = 10.
Number of observations in simulated data = N = 10000.
Number of variables = p = 50.
Number of model parameters = (K - 1) * p = 450.
\end{Soutput}
\end{Schunk}

The next steps setup a \code{formula} object which specifies that  individual specific data must be modeled with alternative specific coefficients and the intercept is excluded from the model.
\begin{Schunk}
\begin{Sinput}
R> vars <- paste("X", 1:50, sep="", collapse=" + ")
R> fm <- formula(paste("response ~ 1|", vars, " - 1 | 1"))
\end{Sinput}
\end{Schunk}
Using this formula and our previously generated \code{data.frame} we run \pkg{mnlogit} to measure its running time (in single threaded mode).
\begin{Schunk}
\begin{Sinput}
R> system.time(fit.mnlogit <- mnlogit(fm, data, "choices"))  
\end{Sinput}
\begin{Soutput}
   user  system elapsed 
  2.200   0.096   2.305 
\end{Soutput}
\end{Schunk}
Likewise we measure running times for \pkg{mlogit} running the same problem with the Newton-Raphson (the default) and the BFGS optimizers.
\begin{Schunk}
\begin{Sinput}
R> mdat <- mlogit.data(data[order(data$indivID), ], "response", shape="long", 
+  alt.var="choices")
R> system.time(fit.mlogit <- mlogit(fm, mdat))   # Newton-Raphson
\end{Sinput}
\begin{Soutput}
   user  system elapsed 
 36.894   5.396  42.431 
\end{Soutput}
\begin{Sinput}
R> system.time(fit.mlogit <- mlogit(fm, mdat, method='bfgs')) 
\end{Sinput}
\begin{Soutput}
   user  system elapsed 
 32.422   8.156  40.731 
\end{Soutput}
\end{Schunk}
Here the first step is necessary to turn the \code{data.frame} object into an \code{mlogit.data} object required by \pkg{mlogit}.
The default stopping conditions for \pkg{mnlogit} and \pkg{mlogit} are exactly the same.
The timing results shown in Table~\ref{table: relative times with mlogit} were obtained in a similar way but with different formulas for each type of problem.
All our tests use the function \code{makeModel} to generate data. 

For comparison with \pkg{nnet} we must make a few modifications: first we turn the data into a format required by \pkg{nnet} and then change the stopping conditons from their default to (roughly) match \pkg{mnlogit} and \pkg{mlogit}.
We set the stopping tolerance so that `reltol' controls convergence and roughly corresponds at termination to `ftol' in these packages.
Note that \pkg{nnet} runs the BFGS optimizer.
\begin{Schunk}
\begin{Sinput}
R> library("nnet")
R> ndat <- data[which(data$response > 0), ]
R> fm.nnet <- paste("choices ~", vars, "- 1")   
R> system.time(fit.nnet <- multinom(fm.nnet, ndat, reltol=1e-12)) 
\end{Sinput}
\begin{Soutput}
# weights:  510 (450 variable)
initial  value 23025.850930 
iter  10 value 22831.705329
iter  20 value 22783.594650
iter  30 value 22777.462151
iter  40 value 22777.307077
iter  50 value 22777.301660
iter  60 value 22777.301615
final  value 22777.301614 
converged
   user  system elapsed 
  3.292   0.000   3.298 
\end{Soutput}
\end{Schunk}
We remind the user that since \pkg{nnet} and \pkg{VGAM} only handle individual specific data, we can't test them on all the classes of problems listed in Table~\ref{table: relative times with mlogit}.
To apply the same timing test to the \code{vglm} function from package \pkg{VGAM}, we first set the stopping condition to match the default condition for \pkg{mnlogit} and \pkg{mlogit} (\code{ftol = 1e-6}).
\begin{Schunk}
\begin{Sinput}
R> library("VGAM")
R> eps <- vglm.control(epsilon = 1e-6)
R> system.time(fit.vglm <- vglm(fm.nnet, data=ndat, multinomial, control=eps))
\end{Sinput}
\begin{Soutput}
   user  system elapsed 
 46.271   1.852  48.298 
\end{Soutput}
\end{Schunk}

Note: The precise times running times reported on compiling this Sweave document depend strongly on the machine, whether other programs are also running simultaneously and the BLAS implementation linked to \proglang{R}.
For reproducible results run on a `quiet' machine (with no other programs running).


\bibliography{mnlogit}
\end{document}